\documentclass[aps,prl,twocolumn,groupedaddress,showpacs]{revtex4}
\usepackage{epsfig}

\def\8{\infty}
\def\oh{\frac{1}{2}}

\def\d{\partial}

\def\undertext#1{\vtop{\hbox{#1}\kern 1pt \hrule}}

\def\be{\begin{equation}}
\def\ee{\end{equation}}
\def\bea{\begin{eqnarray} & &}
\def\eea{\end{eqnarray}}

\def\rf#1{(\ref{#1})}

\def\rf#1{(\ref{#1})}

\def\rfs#1{Eq.~\rf{#1}}

\begin{document}

%\preprint{}

\title{Nonequilibrium dynamics of weakly and strongly paired superconductors}


\author{V. Gurarie}
\affiliation{Department of Physics, University of Colorado,
Boulder CO 80309}

%\thanks{}

\date{\today}

\begin{abstract}
We study small oscillations  of the order parameter in weakly and strongly paired superconductors driven slightly out of equilibrium, in the collisionless approximation. While it was known for quite some time
that the amplitude of the oscillations in a weakly paired superconductor decays as $t^{-1/2}$, we show that in a superconductor sufficiently strongly paired  so that its  fermions form bound states
usually referred to as molecules, these oscillations decay as $t^{-3/2}$. The transition between these two regimes happens when the chemical potential of the superconductor vanishes, thus the behavior of the oscillations can be used to distinguish weakly and strongly paired superconductors. These results are obtained in the mean field approximation which may not be reliable in the crossover region between the strong and weak pairing, so we also obtain identical results within the two-channel model, which can be tuned to be reliable throughout the entire crossover, although it then describes a special type of interactions between the fermions which may be difficult to observe experimentally. Finally, we interpret the result in the strongly paired superconductor as the probability of the molecular decay as a function of time. 

\end{abstract}

\maketitle
The study of quantum quenches, or the evolution of a quantum system when its parameters suddenly change, acquired prominence due to recent studies in Refs.~\cite{Cardy2006,Kollath2007}. Yet a particular case of it, the study of small oscillations of the order parameter in a perturbed superconductor, started more than 30 years ago in Ref.~\cite{Kogan1973}. This paper showed that if a small perturbation is applied to a superconductor leading to a deviation of its gap from its equilibrium value, then this perturbation will evolve as $\cos(2\Delta_0 t)/\sqrt{t}$, where $\Delta_0$ is the equilibrium value of the gap. This work was continued in Ref.~\cite{Spivak2004}. These authors showed that large oscillations of the order parameter in a superconductor do not decay but in fact continue forever (as long as the collisionless approximation remains valid). Subsequently with the help of the exact integrability of this problem \cite{Sierra2004}  it was shown that the exist a critical strength of the perturbation such that if a perturbation beyond that strength is applied, it excites oscillations which continue forever while perturbations below that strength still decay, Ref.~\cite{Barankov2006,Dzero2006}. 

At the same time, advances in atomic physics allow now to create superconductors out of ultracold fermionic atoms whose interactions can be externally controlled. If the interactions are weak, these artificial superconductors behave just like the usual superconductors of condensed matter physics (termed BCS superconductors). But the interactions can also be adjusted to be strong, in which case these superconductors become more akin to a Bose condensate (BEC) of diatomic bosonic molecules. As the interactions are tuned, the superconductor is said to undergo a BCS-BEC crossover (which is accompanied by the chemical potential changing from positive to negative values). The possibility of this crossover was discussed in a number of papers throughout the last four  decades (see Refs.~\cite{Eagles1969,Leggett1980,Nozieres1985}), and it finally was observed a few years ago (see Refs.~\cite{Jin2004,Ketterle2004}). 

The unprecedented control over the interactions that the ultracold gases provide allows to change them quickly thus
easily creating an initial out-of-equilibrium perturbation by a sudden change of their strength.  A natural question which arises in this regard concerns the fate of the oscillations, decaying and persistent, as the superconductor undergoes the BCS-BEC crossover. Refs.~\cite{Andreev2004,Barankov2004} explored the large nondecaying oscillations of a tunable superconductor using the ansatz of Ref.~\cite{Spivak2004}. However, this method was found to break down as the superconductor is tuned to the BEC regime, and it was not possible to tell whether this was an artifact of the technique or an indication that an oscillating solution is indeed not possible in a strongly paired superconductor (see, however, Ref.~\cite{Holland2001a}).

To remedy this situation, in this paper we consider small oscillations of the order parameter in a tunable superconductor. We show that as long as the oscillations remain small, they always decay regardless of the strength of interactions in the superconductor. However, while in the BCS superconductor the amplitude of these oscillations decay as $1/\sqrt{t}$, in the BEC regime they decay as $1/t^{3/ 2}$. The transition between the two regimes happens exactly where the chemical potential $\mu$ is equal to zero. 

The theory we develop is based on the mean field (collisionless) approximation. Thus one can question its validity near the unitary point \cite{Pitaevskii2008} which lies between the BCS and BEC regimes of the superconductors (at positive chemical potential) and is known to go beyond the applicability of the mean field theory \cite{Levinsen2006} (see also Ref.~\cite{Bulgac2009} for recent studies of the dynamics near this point). Therefore, we repeat the calculation in the two channel model \cite{Timmermans1999} describing BCS-BEC crossover with narrow Feshbach resonance \cite{Gurarie2007} where mean field theory is applicable throughout and recover the same result. This shows that an example of the BCS-BEC crossover exists whose order parameter oscillations  unambiguously obey the scenario discussed here. 

In the ``deep BEC" side of the crossover, the superconductor becomes the Bose-Einstein condensate of the diatomic bosonic molecules. In this regime, the result $1/t^{3/ 2}$ has a very simple interpretation. Indeed, as we will see in this paper, the order parameter of such superconductor is proportional to the probability of finding two fermions of opposite spins at the same position in space. As the superconductor is perturbed, the wave function of a pair of fermions no longer coincides with their bound state, but rather consists of the linear combination of a bound and excited states. The part of the wave function in the excited states moves off to infinity so that the probability amplitude to find two particles in the same spot decreases as $1/t^{3/2}$ reaching a limiting
finite value at large $t$ which in turn follows from the behavior of a three dimensional propagator of a free particle. In fact, in this deep BEC regime this is true not only with a small but also with a relatively large initial perturbation (the precise criteria are developed below), in which case this picture makes it clear that the order parameter decays to a value smaller than the equilibrium value, just like
in the BCS regime with certain perturbations \cite{Dzero2006}. 
%Combined with the fact that no persistent oscillatory solutions were found in the BEC regime so far, this makes the possibility of the existence of such non-decaying oscillations in the BEC regime doubtful. \
However, to fully analyze the case of large perturbations to the superconductor and to see whether nondecaying solutions are possible in the BEC regime, one needs to take advantage of the integrability of the equations of motion, something which was done successfully for the weakly paired superconductor and whose strongly paired superconductor applications are left for future work. 

In what follows, we describe the derivation of these results. Consider spin-1/2 fermions interacting via a short range attractive $s$-wave interaction of strength $\lambda$. Within the mean field approximation, these fermions can be studied on a strictly classical level. We accomplish this by introducing the Cooper pair number $n_p = \left< \hat a^\dagger_{\uparrow p} a_{\uparrow p} + \hat a^\dagger_{\downarrow p} a_{\downarrow p} \right>/2$. The evolution of this number obeys the following classical Hamiltonian, a classical version of the Anderson-Richardson (or reduced BCS) Hamiltonian describing these kind of fermions,
\be  \label{eq:ham}
H = 2 \epsilon_p n_p - \frac{\lambda}{V} \sum_{p,q} \sqrt{n_p (1-n_p) n_q (1-n_q)} \cos(\phi_p-\phi_q),
\ee
where $\epsilon_p=p^2/(2m)$ is the free fermions dispersion, $\phi_p$ is the phase variable canonically conjugate to $n_p$, and $V$ is the space volume. The equations of motion of this Hamiltonian have the following solution, corresponding to a stationary superconductor
\be  \label{eq:gap}
\phi_p^0 = -2\mu t, n_p^0=\oh \left( 1- \frac{\xi_p}{E_p} \right),
\ee
where as always $\mu$ is the chemical potential, $t$ is time, $\xi_p=\epsilon_p-\mu$, $E_p=\sqrt{\xi_p^2+\Delta_0^2}$, and finally $\Delta_0$ is the equilibrium gap of the superconductor which can be found from
the gap equation
$1 = \frac{\lambda}{2V} \sum_p \frac 1 {E_p}.$
Now consider the initial conditions for the motion described by \rfs{eq:ham} consisting of a small perturbation to the stationary solution \rfs{eq:gap}. In fact, it is most physical to take as initial conditions \rfs{eq:gap} for a slightly different value of $\Delta= \Delta_0+\delta \Delta_0$ as well as the slightly different value of $\mu$ (corresponding to a superconductor whose interactions were slightly perturbed in the initial moment of time). Then it is straightforward to see, with the help of the particle conservation condition $\sum_p \delta n_p=0$, that the initial conditions read
\be \label{eq:ini} n_p = n_p^0+ \delta n_p^0, \ \delta n_p^0 = \left( \xi_p - \frac{\tilde f_0}{f_0} \right) \frac{\Delta_0 \delta \Delta_0}{2 E_p^3},
\ee
where $f_0$ and $\tilde f_0$ are defined below in \rfs{eq:f0}. 
We then expand the Hamiltonian \rfs{eq:ham} about the stationary solution to obtain the quadratic Hamiltonian for the deviations $\delta n_p$, $\delta \phi_p$. Subsequently we construct the solution to the equations of motion of this  Hamiltonian by using the method of Green's functions. The expanded Hamiltonian takes the form
\be  \delta H = \oh \sum_{p,q} \left[ \delta \phi_p \Phi_{pq} \delta \phi_q + \delta n_p K_{pq} \delta n_q \right],
\ee
where 
\be \Phi_{pq} = \frac{\Delta_0^2}{E_p} \delta_{pq} - \frac{\lambda}{2V} \frac{\Delta^2}{E_p E_q}, K_{pq} = \frac{4 E_p^3}{\Delta_0^2} \delta_{pq} - \frac{2 \lambda \xi_p \xi_q}{V \Delta_0^2}.
\ee
This represents a collection of harmonic oscillators, labelled by the index $p$. 
Now it is possible to construct a retarded Green's function corresponding to these oscillators, with the end result
\be \label{eq:sol} \delta n_p(t) =i \sum_q \int_{-\infty+i0}^{\infty+i0} \frac{  \Omega d\Omega}{2\pi} \, G_{pq}(\Omega) \, e^{-2 i \Omega t} \, \delta n_q^0,
\ee
where
\begin{eqnarray}  \label{eq:green}
&& G_{pq}(\Omega)  =  \frac{\delta_{pq}}{\Omega^2-E_p^2} + \\ &&
\frac{\lambda}{2V} \frac{f \left( \Omega^2 \xi_p \xi_q + E_q^2 \left(\Omega^2-\Delta^2_0 \right) \right)-\tilde f \left( \xi_p E_q^2+ \xi_q \Omega^2 \right)}{\Omega^2 E_p \left(\Omega^2-E_p^2 \right) \left(\Omega^2-E_q^2 \right)
\left[ \tilde f^2 + f^2 \left(\Delta_0^2 - \Omega^2 \right) \right]}. \nonumber
\end{eqnarray}
Here we introduced the functions
\be \label{eq:f}  f=\frac{\lambda}{2V} \sum_p \frac{1}{E_p \left(\Omega^2-E_p^2\right)} ,  \ \tilde f = \frac{\lambda}{2V} \sum_p \frac{\xi_p}{E_p \left(\Omega^2-E_p^2\right)}.
\ee
In what follows we will also need these same functions evaluated at $\Omega=0$, or
\be  f_0 =-\frac{\lambda}{2V} \sum_p \frac{1}{E_p^3}, \ \tilde f_0 = -\frac{\lambda}{2V} \sum_p \frac{\xi_p}{E_p^3}. \label{eq:f0}
\ee
%Note that these same functions were introduced by the authors of Ref.~\cite{Kogan1973}, however they only considered $f$ and $f_0$ as $\tilde f$ and $\tilde f_0$ vanish in the weakly paired superconductor, the only
%regime considered in that paper. 
The derivation of \rfs{eq:green} is technical and not particularly instructive, so it will be published elsewhere. 

\begin{figure}[h]
\includegraphics[height=1 in]{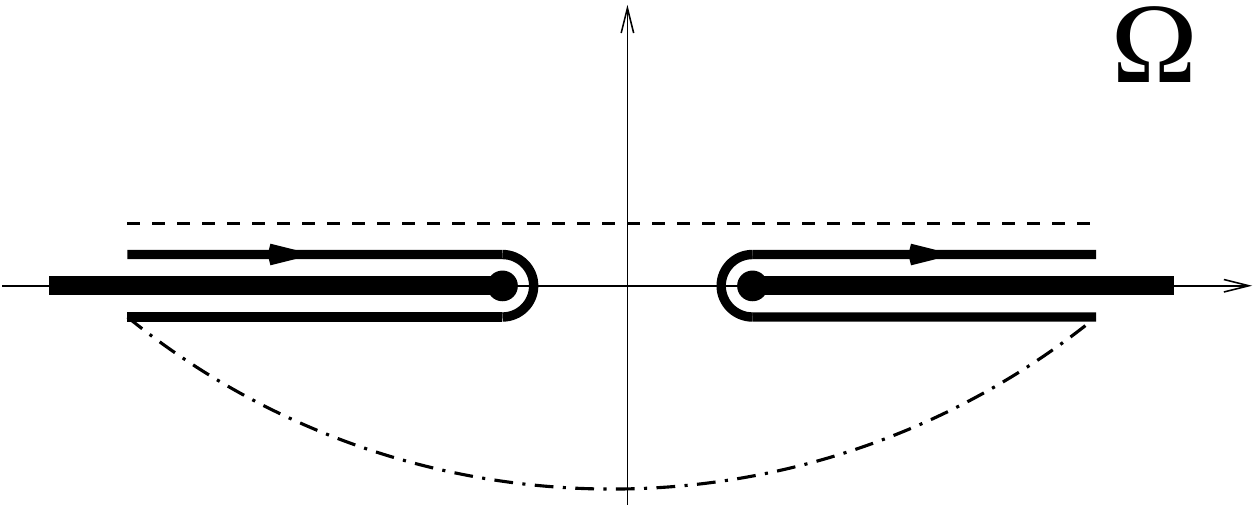} \caption{The contour of integration in \rfs{eq:flucd} in the complex plane of $\Omega$. Dashed line shows the initial contour, while solid black line, together with a dashed-dotted semicircle, is the deformed contour. The dashed-dotted part of the new contour can be deformed away to infinity and does not contribute. }
 \label{Fig3}
\end{figure}

Armed by the explicit expression for the fluctuations $\delta n_p(t)$ we can now discuss how to calculate the time dependent gap function. 
It is easiest to study the square of its absolute value,
\be \left| \Delta(t) \right|^2 = \frac{\lambda^2}{V^2} \sum_{p,q} \sqrt{n_p (1-n_p) n_q (1-n_q)} \cos(\phi_p-\phi_q).
\ee
In turn, this quantity can be decomposed into the sum of $\Delta_0^2$, the square of the unperturbed gap, and the perturbation $\delta \left| \Delta \right|^2$, given by
\be
\delta \left| \Delta(t) \right|^2 = \frac{2\lambda}{V} \sum_p \xi_p \, \delta n_p(t).
\ee
We combine this with the initial conditions \rfs{eq:ini} as well as with Eqs.~\rf{eq:sol} and \rf{eq:green} to find that
\begin{eqnarray} \label{eq:flucd} &&  \delta \left| \Delta(t) \right|^2 =  \frac{ i  \Delta_0 \delta \Delta_0}{\pi}  \int_{-\infty+i0}^{\infty +i0} \Omega d\Omega \, e^{-2 i \Omega t}  \times \\ && \frac{\tilde f^2 + f^2 \left(\Delta_0^2-\Omega^2 \right) - f  f_0 \Delta_0^2 - f \tilde f_0^2/f_0 }{\Omega^2
\left[ \tilde f^2 + f^2 \left( \Delta^2_0-\Omega^2 \right) \right]}. \nonumber
\end{eqnarray}
This equation represents the main result of this paper. White reducing to the main result of Ref.~\cite{Kogan1973} at weak pairing, it represents the generalization of their result to a superconductor of an arbitrary pairing strength. The integration here goes over the straight line slightly above the real axis, as shown on Fig.~\ref{Fig3} with a dashed line.
We expect the functions $f$ and $\tilde f$ to have singularities on the real axis where $\left| \Omega \right| $ hits the minimum of $E_p$ and cuts at larger $\left| \Omega \right|$. It may seem that there is also a singularity at $\Omega=0$, but it is actually fictitious as the numerator of \rfs{eq:flucd} vanishes at $\Omega \rightarrow 0$. At $t<0$ the contour can be closed in the upper half plane resulting in zero for the integral as expected. At positive time $t>0$ we can deform the contour as shown on Fig~\ref{Fig3} by a solid line with arrows. Then at large positive times, the main contribution to the integral comes from the vicinity of the singular points, or the turning points of the new contour. 

To evaluate $f$ and $\tilde f$, we replace summation over $p$ in \rfs{eq:f} by the integration over $V d^3p/(2\pi)^3$. 
In the BCS regime, it is standard to pass to the integration over the variable $\xi=\epsilon_p-\mu$ so that $d^3p/(2\pi)^3
\approx  \nu d\xi$, where $\nu=\sqrt{m^3 \mu} /(\pi^2\sqrt{2})$ is the density of states, and then extend the integration over $\xi$ all the way to the entire real axis.  Then $\tilde f=0$  as its integral is antisymmetric in $\xi$. At the same time, $f$ is given by
\be f = i \frac{\nu \lambda}{\Omega \sqrt{\Delta_0^2-\Omega^2} } \ln \left[ \sqrt{1-\frac{\Omega^2}{\Delta_0^2}} + i \frac{\Omega}{\Delta_0} \right].
\ee
It is then easy to see that the integrand in \rfs{eq:flucd} has a singularity as $\Omega$ approaches $\Delta_0$ which goes as $1/\sqrt{\Omega-\Delta_0}$ in agreement with the discussion above ($\Delta_0$ coincides with the minimum of $E_p$). Introducing the variable $s=\Omega-\Delta_0$ and taking into account that at $t \gg 1/\Delta_0$ only small $s$ contributes to the integral, we find  after adding both left and right contours (which are complex conjugate of each other)
\be \label{eq:bcs}
\delta \left| \Delta(t) \right|^2 \sim {\rm Re}~ e^{-2 i \Delta_0 t} \int_0^\infty \frac{ds e^{-i s t}}{\sqrt{s}} \sim \frac{\cos(2 \Delta_0 t)}{\sqrt{t}}.
\ee
Alternatively, in the BEC regime where $\mu<0$, the integration measure in \rfs{eq:f} must be kept as $V d^3p/(2\pi)^3$ and both functions $f$ and $\tilde f$ are not zero. The minimum of $E_p$  is equal to $E_{\rm min}=\sqrt{ \mu^2+ \Delta_0^2} $. The functions $f$ and $\tilde f$ are now finite when  $\left|\Omega\right|$ reaches $E_{\rm min}$, but they are not regular at that point. Generally they behave as $f \sim {\rm const} +i \sqrt{\Omega-E_{\rm min} }$. The origin of such a drastically different behavior lies in the fact that the minimum of $E_p$ is reached when $p=0$. Then, the potential divergence in the integral at $\Omega=E_{\rm min}$ is removed by the $p^2$ coming from the measure of integration $p^2 dp$. All this can be checked explicitly in the deep BEC regime where $\left| \mu \right| \gg \Delta_0$  and the analytic expressions behaving precisely in this way can be obtained by direct integration.%Then the integrals can again be computed analytically, with the result

%\begin{eqnarray} f&=& \frac{\lambda m^{\frac 3 2} \sqrt{\left| \mu \right|}}{\sqrt{2}\pi \Omega^2} \left[-1 + \oh \left( \sqrt{1+\frac{\Omega}{\mu}} + \sqrt{1-\frac{\Omega}{\mu}} \right) \right],\cr
%\tilde f &=& -\frac{\lambda m^{\frac 3 2}}{\sqrt{2} \pi \sqrt{\left| \mu \right|}} \left[ \sqrt{1+\frac{\Omega}{\mu}} + \sqrt{1-\frac{\Omega}{\mu}} \right]^{-1}.
%\end{eqnarray}
%These functions have the behavior discussed above. 

Now if we substitute this into \rfs{eq:flucd} we find that
the constant piece is single valued as $\Omega$ goes from the upper to lower branch of the contour on Fig.~\ref{Fig3} and therefore its contribution from each of the contours cancels. So it is the square root part which contributes. This gives at large times $t \gg 1/E_{\rm min}$
\be \label{eq:bec} \delta \left| \Delta \right|^2 \sim {\rm Re}~ e^{-2 i t E_{\rm min}} \int ds \, e^{-i s t} \sqrt{s} \sim \frac{\cos(2 t E_{\rm min})}{t^{\frac 3 2}}.
\ee
As we see, the transition from the behavior \rfs{eq:bcs} to \rfs{eq:bec} occurs exactly at $\mu=0$ as this is the point where the minimum of the spectrum $E_p$ shifts to $p=0$. 

One can worry if this derivation really works for a superconductor in the vicinity of the unitary point where the mean field theory breaks down. To have a controllable theory, we repeated this calculation for the two channel model, based on the equations of motion introduced in Ref.~\cite{Andreev2004}. That model has an additional coupling $g$ such that if $g^2 m^2/n^{1/3} \gg 1$ (here $n$ is the density), then the two channel model is equivalent to \rfs{eq:ham} considered earlier in this paper. In the opposite limit, the two channel model still undergoes a crossover but with the mean field theory applicable throughout. 
The answer for the two channel model can be worked out using the same methods as the ones described here. It looks almost identical to \rfs{eq:flucd}, except in all occurences of $\tilde f$ and $\tilde f_0$ one needs to substitute $\tilde f \rightarrow \tilde f -
{2 \lambda}/{g^2}$,  $\tilde f_0 \rightarrow \tilde f_0 -
{2 \lambda}/{g^2}$. This replacement does not change any of the arguments presented here for the one channel model, thus all the conclusions remain valid.

Finally, we observe that in the deep BEC regime all $n_p^0 \ll 1$. Then the Hamiltonian \rfs{eq:ham} can be significantly simplified, reducing in this limit to the Schr\"odinger equation of one pair of fermions in a delta-function potential, as discussed at length in Ref.~\cite{Nozieres1985}. Replacing $1-n_p \approx 1$, and introducing the pair wave function $\psi_p =\sqrt{ n_p} \, e^{i \phi_p}$, 
we find  $H = 2 \epsilon_p \left| \psi_p \right|^2- \frac{\lambda}{V} \left| \sum_p \psi_p \right|^2$ leading to
\be \label{eq:sch}  i \d_t \psi = 2\epsilon_p \psi_p-
\frac \lambda V \sum_q \psi_q.
\ee
The ``gap function" in this language is $\Delta = \lambda \sum_p \psi_p/V$ which has an obvious meaning of the probability of finding  two opposite-spin fermions at the same point in space. 
%As discussed earlier in this paper, this probability is expected to decay as $1/t^{\frac 3 2}$ simply because this is the escape rate in 3 dimensional space. 
We can solve the linear equation \rf{eq:sch} with arbitrary initial
conditions directly. The solution for the gap function reads
\begin{eqnarray} \label{eq:ansbec} \Delta(t) &=& \frac{\lambda}{V} \sum_p \int_{-\infty+i0}^{\infty+i0} d\Omega\, e^{-i \Omega t} \frac{1}{1+\Pi} \frac{\psi_p^0}{\Omega-2 \epsilon_p}, \cr \Pi &=& \frac{\lambda}{V}\sum_p \frac{1}{\Omega-2\epsilon_p} = 
{\rm const} +\frac{ \lambda m^{\frac 3 2} \sqrt{-\Omega}}{4 \pi}
\end{eqnarray} with  $\psi_p^0$ being the initial value of $\psi_p$. To compute the integral over $\Omega$, we deform the contour of integration in a way similar to that shown on Fig.~\ref{Fig3}. Unlike \rfs{eq:flucd},
the integrand here has a simple pole at a negative value of $\Omega=\Omega_b$ where $\Pi(\Omega_b)=-1$, but similarly to \rfs{eq:flucd} in the BEC regime, it has cut at $\Omega>0$ where it behaves as ${\rm const}+i\sqrt{\Omega}$. The pole corresponds to the bound state and results in a contribution to $\Delta(t)$ corresponding to the bound molecule (that is, the equilibrium solution). The cut produces the decaying behavior going precisely as $1/t^{3/2}$, thus overall $\left|\Delta(t) \right|^2 \approx {\rm const} + \cos(\Omega_b t)/t^{3/2}$ (for comparison with \rfs{eq:bec} observe that $\Omega_b=2\mu$ in this regime). However now we see that this decay is correct for arbitrary initial conditions, as long as the applicability conditions $\left| \psi_p^0 \right| \ll 1$ are satisfied. The value $\left|\Delta(t) \right|^2$ decays to is smaller than the equilibrium value of the gap function, since it is reduced by the overlap, smaller than 1, of the initial state and the bound state. 
%The only non-smooth $\psi_p^0$ which can realistically be produced as initial state of fermions would correspond to a free zero temperature %Fermi-Dirac distribution, and taken naively these can led to nondecaying oscillations. 
The only  initial distributions out of reach of this method are the ones where $\left|\psi_p^0 \right|$ are no longer much smaller than 1.

In conclusion, we showed that the amplitude of the small oscillations of the order parameter in a superconductor decays as $1/t^{1/2}$ in the BCS ($\mu>0$) and as $1/t^{3/2}$ in the BEC ($\mu<0$) regimes. Applicability of this result to the Fermi gas close to the unitary point remains an open question. It would also be very interesting to see if, with modern experimental techniques, these oscillations can be detected in real superconductors and whether this theory can be generalized to superconductors with  the $d$-wave symmetry of the gap. 
%This is definitely valid for narrow resonance BCS-BEC crossover. On the other hand, the applicability of these results for the Fermi gas close to the unitary point remains an open question. The assumption of smallness of initial perturbation can be relaxed in the deep BEC regime leading to the same result, however to see the absence or presence of persistent oscillations when one starts from a free Fermi gas distribution one probably needs to go beyond the methods developed in this paper. 

This work was supported by the NSF grant DMR-0449521. The author is grateful to M. Hermele, L. Radzihovsky, as well as R. Barankov and Chen Zhang, for many discussions. 

\bibliography{paper}

\end{document}